\documentclass[twocolumn,usenames,dvipsnames]{aastex62}
\usepackage{amsmath}
\usepackage{apjfonts}
\usepackage{amssymb}
\usepackage{xspace}
\usepackage{hyperref}
\usepackage{natbib}
\usepackage{multirow}
\usepackage{graphicx}
\usepackage{epstopdf}
\usepackage{epsf}
\usepackage{subfigure}
\usepackage{xcolor}
\usepackage{rotating}
\usepackage{mathrsfs}
\usepackage{slashbox}
\usepackage{multirow}
\usepackage{lipsum}
\usepackage[normalem]{ulem}
\usepackage{diagbox}

\def\sigL{\sigma_{L}}
\def\sqrtJD{\sqrt{\Delta_r \Delta_z}}
\def\CHIMP{h^{-1} \rm Mpc}
\def\massu{h^{-1} \rm M_\odot}

\usepackage{lineno}

\begin{document}

\title{Approximating Density Probability Distribution Functions Across Cosmologies}

\author{Huanqing Chen}
\affiliation{Department of Astronomy \& Astrophysics; 
The University of Chicago; 
Chicago, IL 60637, USA}

\author{Nickolay Y.\ Gnedin}
\affiliation{Theoretical Physics Department; 
Fermi National Accelerator Laboratory;
Batavia, IL 60510, USA}
\affiliation{Kavli Institute for Cosmological Physics;
The University of Chicago;
Chicago, IL 60637, USA}
\affiliation{Department of Astronomy \& Astrophysics; 
The University of Chicago; 
Chicago, IL 60637, USA}

\author{Philip Mansfield}
\affiliation{Kavli Institute for Particle Astrophysics and Cosmology; Stanford University, Stanford, CA 94305, USA}

\correspondingauthor{Huanqing Chen}
\email{hqchen@uchicago.edu}

\begin{abstract}
Using a suite of self-similar cosmological simulations, we measure the probability distribution functions (PDFs) of real-space density, redshift-space density, and their geometric mean. We find that the real-space density PDF is well-described by a function of two parameters: $n_s$, the spectral slope, and $\sigL$, the linear rms density fluctuation. For redshift-space density and the geometric mean of real- and redshift-space densities, we introduce a third parameter, $s_L={\sqrt{\langle(dv^L_{\rm pec}/dr)^2\rangle}}/{H}$. We find that density PDFs for the LCDM cosmology is also well-parameterized by these three parameters. As a result, we are able to use a suite of self-similar cosmological simulations to approximate density PDFs for a range of cosmologies. We make the density PDFs publicly available and provide an analytical fitting formula for them.

\end{abstract}

\section{Introduction}

The prevailing inflation theory predicts that the initial perturbations in the early universe are tiny, Gaussian, and nearly scale-invariant \citep[e.g.][]{mo2010}. As a result, the 1-point density PDF, $\mathcal{P}(\delta)$, has an almost perfectly Gaussian form in the early universe. The subsequent evolution of the matter field under gravity distorts this distribution from its initial shape. Dark energy, dark matter, baryons, and relativistic species all impact the evolution of the matter density field. Therefore, studying the distribution of matter helps us to constrain cosmology.

The probability distribution function (PDF) of matter density in the universe is a simple but important statistical function for cosmology. It was one of the first statistical functions measured by extra-galactic surveys \citep[e.g.,][]{hubble1934, soneira1978, Efstathiou1990}. The density PDF is essentially independent from the two-point correlation function, which has been extensively studied during the digital galaxy survey era. Hence, the density PDF contains information about correlation functions of order higher than two. For example, \citep{uhlemann2020} showed that by combining the density PDF with other statistical tools much tighter constraints can be made on the sum of neutrino masses, demonstrating the great potential of using the density PDF in constraining cosmology.

The density PDF has been studied extensively using perturbation theory and excusion set models \citep[e.g.,][]{juszkiewicz1993, bernardeau1994,sheth1998,lam2008}. However, such prescriptions only work well in the linear or mildly non-linear regime.
More recent studies have used simulations to obtain density PDFs and provide fitting formulae \citep[e.g.,][and references therein]{bernardeau2002,shin2017, klypin2018,ivanov2019} 

In practice, the density probability distribution function at low redshifts is usually measured by galaxy counts-in-cell statistics. However, because galaxies have peculiar velocity on top of the Hubble flow, their real positions can not be measured. To directly compare such an observed PDF, one needs to have a model for the redshift-space density, $\Delta_z$. There have been a few studies focusing on the redshift-space density PDF \citep[e.g.,][]{watts2001}.
At very high redshifts, especially during cosmic reionization, galaxy-based counts-in-cell statistics are very challenging because it is hard to detect enough galaxies and to estimate the pureness and completeness of the sample.
Recently, \citet{chen2021b} found that the gas density can be recovered in quasar proximity zones at $z\sim 6$, providing a way to measure the density PDF during cosmic reionization. The recovered density, however, is not a redshift-space density (as can be expected from a measurement of an observed spectrum), but rather the geometric mean $\sqrtJD$ of the real-space density $\Delta_r$ and the redshift-space density $\Delta_z$.
For that reason, we also consider a PDF model for such a geometric mean density in this paper.

It is hard to find a precise PDF model for arbitrary cosmologies. However, it is possible to obtain approximate functions. In this case, self-similar simulations are very useful. These simulations are run with a power-law initial power spectrum, and contain only dark matter without dark energy or baryons (i.e. $\Omega_m \equiv 1$). In such a simple cosmology, we can gain useful insights into the shape of the density PDF \citep{colombi1997}. Because there is no scale dependence in such a cosmology, one can measure the statistic at one redshift and apply it to all other redshifts.
The goal of this paper is two-fold. First, we measure PDFs for real-space density, redshift-space density and $\sqrtJD$ in a suite of N-body simulations with self-similar cosmologies. We introduce a way to parameterize them and provide fitting formulae for a range of parameters. Second, we show that the density PDFs inferred from self-similar cosmologies are good approximations for those found in a WMAP-like $\Lambda$CDM cosmology, using the value of the power spectrum slope at the scale corresponding to the smoothing kernel. Therefore, this method allows us to calculate the density PDFs of a wide range of cosmologies with just a suite of self-similar simulations.

\section{Parameterizing Density PDFs}

In practice, the density PDF is measured with a specific smoothing kernel. In this study, we study the density fields smoothed by a 3D Gaussian kernel of size $R$,

\begin{equation}
{W}_{\rm G3D}(\textbf{x},R) = \frac{1}{(2 \pi)^{3/2} R^3} \exp\left[-\frac{|\textbf{x}|^2}{2 R^2}\right].
\end{equation}

Let us first consider a self-similar cosmology, i.e., one with the power-law initial power spectrum and $\Omega_m=1$. In this section, we show how one can parameterize the density PDFs for such a cosmology, and in the next section, we demonstrate that this parameterization works well for both self-similar cosmology and for a commonly-used $\Lambda$CDM cosmology.

\subsection{The real-space Density PDF}
\label{sec:param}

We parameterize the real-space density PDF for the self-similar cosmology with two parameters, the slope of the initial power spectrum, $n_s$, and the linear rms density fluctuation of the smoothed density field, $\sigL$:

\begin{equation} \label{eq1}
\begin{split}
\sigL^2 (R,z) & \equiv \frac{1}{(2\pi)^3} \int P(k,z) |{\tilde{W}}(kR)|^2\, d^3 k \\
 & = \frac{1}{2\pi^2}\int k^2 P_L(k,z){\tilde{W}}^2(kR)\, dk.
\end{split}
\end{equation}

Here $\tilde{W}$ is the smoothing kernel in Fourier space. In this paper, we use a 3D Gaussian kernel: 
\begin{equation}
\tilde{W}_{\rm G3D}(kR) = \exp\left[-\frac{(kR)^2}{2}\right].
\end{equation}
    Note that $\sigL$ is a function of both the smoothing scale $R$ and the redshfit $z$. However, for a self-similar cosmology, \citet{bernardeau1994}  showed that the density cumulants for a top-hat-smoothed field only depend on $\sigL$ while in the linear and quasi-linear regimes. In other words, for the same value of $\sigL$, density PDFs at different redshifts should be the same. We show that this is also true for Gaussian smoothing kernels in Section \ref{sec:real_density_pdf}.

\subsection{The redshift-space density and $\sqrt{\Delta_r \Delta_z}$   PDFs}\label{sec:param_zsPDF}

Peculiar velocities distort the redshift-space density field. The redshift-space density is related to the real-space density by
\begin{equation} \label{Eq:tradDz}
    \Delta_{z}= \Delta_{r} \left| H\frac{dr_{\bf n}}{du_{\bf n}} \right|=\Delta_{r} \left| \frac{1}{1+(dv_{\bf n}/dr_{\bf n})/H} \right|,
\end{equation}
where $H$ is the Hubble parameter, $u_{\bf n} \equiv {\bf u\cdot n}$ and $v_{\bf n} \equiv \bf v\cdot n$ are the total and the peculiar velocity along the line of sight respectively, and $r_{\bf n}$ is the proper distance along the direction $\bf n$. From this equation,
a natural choice to parameterize the redshift-space density PDF is to introduce a third (dimensionless) parameter $s$, which we shall call the distortion parameter:

\begin{equation}\label{eq:s}
s=\frac{\sqrt{\langle(dv_{\bf n}/dr_{\bf n})^2\rangle}}{H}.
\end{equation}

In the linear regime, peculiar velocity and density are related through
\begin{equation}\label{eq:sL}
{v_{{\bf n}, \bf k}^{\rm L}} = \frac{i a k_{\bf n}}{k^2}\frac{d \delta_{\bf k}}{dt}=\frac{i {\bf kn}}{k^2}Ha\delta_{\bf k}^L f(z), 
\end{equation}
and
\begin{equation}
\left(\frac{dv_{\bf n}}{dr_{\bf n}}\right)^L_{\bf k} = \frac{({\bf kn})^2}{k^2} H {\delta}_{\bf k} f(z),
\end{equation}
where 
\begin{equation}\label{eq:fz}
f(z) \equiv -\frac{d \ln D(z)}{d\ln (1+z)}
\end{equation}
and $D(z)$ is the linear growth rate at redshift $z$. As a result, 
\begin{equation}
\begin{split}
    s_L &=\frac{\sqrt{\langle(dv^L_{\bf n}/dr_{\bf n})^2\rangle}}{H} \\
        &=f(z)\sqrt{\frac{1}{(2\pi)^3}\int d^3{\bf k} \, \frac{({\bf kn})^4}{k^4} P_L(k,z)W^2(kR)} \\
        &=\frac{f(z)}{\sqrt{5}}\sigL.
\end{split}
\end{equation}

Therefore, in the linear regime, the distortion parameter is proportional to $f(z)$.
For a self-similar cosmology,  $f(z) \equiv 1$, and $s_L$ is a redundant parameter. Therefore, the redshift-space density will also be independent of redshift, as will be demonstrated in the next section. However, for a general cosmology $f$ is not necessarily equal to 1.

The geometric mean of the real- and redshift-space densities $\sqrtJD$  is related to the real-space density via
\begin{equation} \label{Eq:sqrtJden}
    \sqrtJD= \Delta_{r} \sqrt{\left| \frac{H dr_{\bf n}}{du_{\bf n}} \right|}=\Delta_{r} \sqrt{\left| \frac{1}{1+(dv_{\bf n}/dr_{\bf n})/H} \right|}.
\end{equation}
Hence, the PDF of $\sqrtJD$ also depends on $s_L$ and $f(z)$.

\section{Measuring PDFs from Simulations}

\subsection{Simulations with Self-similar Cosmologies}

We first analyze a suite of scale-free simulations. A full description of these simulations can be found in \citet{diemer2015}. Here we provide some key information about them and refer the reader to the original paper for more details. The simulation suite consists of four simulations with power-law initial power spectra. These spectra have indices $n_s=-1.0, -1.5, -2.0,$ and $-2.5$. The initial conditions are generated with 2LPT{\footnotesize IC} \citep{crocce2006}, using the 2$^{\rm nd}$-order perturbation theory. All four simulations have box sizes of $100\ \CHIMP$ per side, and are run using the \textsc{Gadget-2} code with  $N^3=1024^3$ particles. The mass resolution is $m_p=2.6\times 10^8\ \massu$.

For a grid of $\sigL$ and $n_s$ values listed in Table \ref{tab:param_grid}, we calculate the density PDF as follows. First, we deposit particles onto a $1024^3$ grid via the cloud-in-cell method to create density fields, $\rho$, and $\rho v_i$ fields. Here, $i$ indexes over spatial dimensions. Then for each $\sigL$ and $n_s$ combination, we smooth the data cubes with a 3D Gaussian kernel using the values for $R$ listed in Table \ref{tab:param_grid} using Fast Fourier Transform. We directly measure the real-space density PDFs from the smoothed data cubes. To measure the redshift-space density PDFs, we draw $64\times 64$ uniformly spaced lines parallel to each side of the smoothed data cube. We calculate the density weighted velocity and then obtain the redshift-space density, from which we measure both the redshift-space density and $\sqrtJD$.

\begin{table}[]
\begin{tabular}{|l|l|l|l|l|l|l|}
\hline
\multirow{2}{*}{$n_s$}                         & \multirow{2}{*}{$z$} & \multicolumn{5}{c|}{ $R \  [\CHIMP]$}                                                                                                                        \\ \cline{3-7} 
                                             &                      & $\sigL=0.4$                  & $\sigL=0.6$                  & $\sigL=0.8$                 & $\sigL=1.0$                 & $\sigL=1.2$                 \\ \hline
\multicolumn{1}{|c|}{\multirow{3}{*}{-1.0}} & 9                    & 0.76                          & -  & - &         -                     &      -                        \\ \cline{2-7} 
\multicolumn{1}{|c|}{}                       & 4                    & 1.54                          & 1.03                          & 0.77                         & 0.62                         & - \\ \cline{2-7} 
\multicolumn{1}{|c|}{}                       & 2                    & -            & 1.72                          & 1.29                         & 1.03                         & 0.86                         \\ \hline
\multirow{3}{*}{-1.5}                       & 5                    & 0.90                          & -  &     -    &      -  &    -   \\ \cline{2-7} 
                                             & 2                    & 2.28                          & 1.33                          & 0.91                         & 0.67                         & - \\ \cline{2-7} 
                                             & 1                    &  -                   & 2.26                          & 1.54                         & 1.14                         & 0.90                         \\ \hline
\multirow{4}{*}{-2.0}                       & 4                    & 0.63                          & -  &               -          &                -              &   -                           \\ \cline{2-7} 
                                             & 2                    & 1.67                          & 0.74                          & -   &                 -             &         -                     \\ \cline{2-7} 
                                             & 1                    & -                   & 1.74                          & 0.98                         & 0.63                         & - \\ \cline{2-7} 
                                             & 0.5                  & -  & 3.16                          & 1.78                         & 1.14                         & 0.79                         \\ \hline
\multirow{4}{*}{-2.5}                       & 2                    & 0.73                          & -  &    -   &  -    & -                \\ \cline{2-7} 
                                             & 1                    & -  & 0.80                          & - &       -                      &   -                           \\ \cline{2-7} 
                                             & 0.5                  & -& -           & 0.84                         & - &  -                            \\ \cline{2-7} 
                                             & 0                    &   -                        & - & -     & 1.50           & 0.72                         \\ \hline
\end{tabular}
\caption{The $n_s$ and $\sigma_L$ combinations used in this paper, along with their corresponding redshifts, $z$, and smoothing scales, $R$. We restrict $R$ to the $0.6\sim\ 3 \CHIMP $ range to avoid discretization effects at the low-$R$ limit and to ensure that there are still sufficiently many independent samples at the high-$R$ limit.}
\label{tab:param_grid}
\end{table}

\subsubsection{Real-Space Density PDFs}
\label{sec:real_density_pdf}

\begin{figure}
    \centering
    \includegraphics[width=0.5\textwidth]{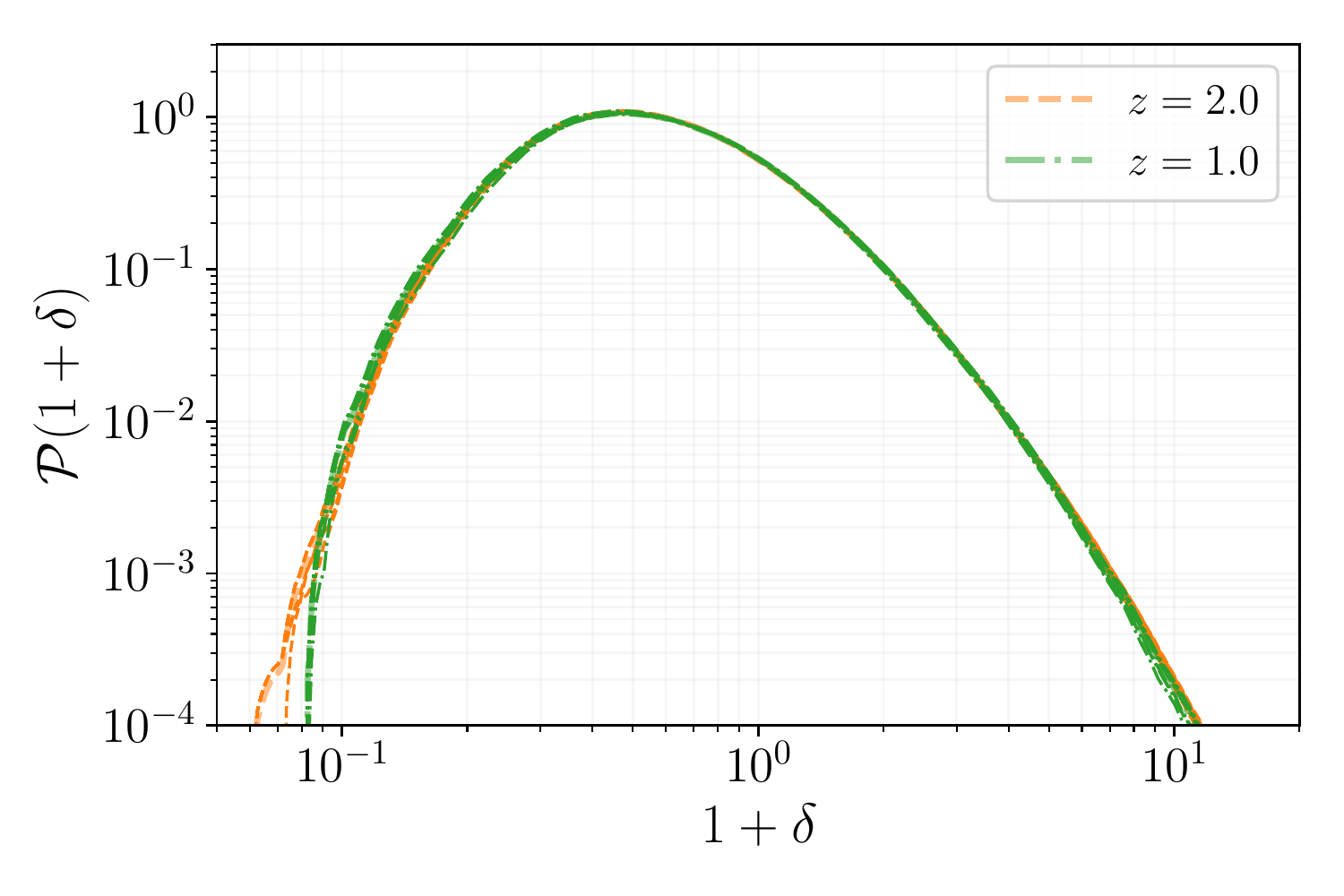}
    \caption{Real-space density PDF for  $\sigL=0.8$ and $n_s=-1.5$ at two different redshifts $z=2$ (orange) and $z=1$ (green).}
    \label{fig:rPDF_diff_z}
\end{figure}

As mentioned in Section \ref {sec:param}, the real-space density PDF should not explicitly depend on redshift for self-similar cosmologies, but only on $n_s$ and $\sigL$.
To verify this, in Figure \ref{fig:rPDF_diff_z} we show the real-space density PDFs for $n_s=-1.5$ and $\sigL=0.8$ at two different redshifts $z=2$ (orange) and $z=1$ (green).
To estimate the sample variance, we divide simulation boxes into $8$ octants and iteratively exclude one octant at a time to calculate $8$ ``jackknifed'' PDFs, which are overlaid in Figure \ref{fig:rPDF_diff_z} with thinner but more opaque lines of the same color. As expected, the real-space PDFs are almost identical at these two different redshifts, except at the lowest densities, where numerical effects of grid discretization become significant.

\begin{figure}
    \centering
    \includegraphics[width=0.5\textwidth]{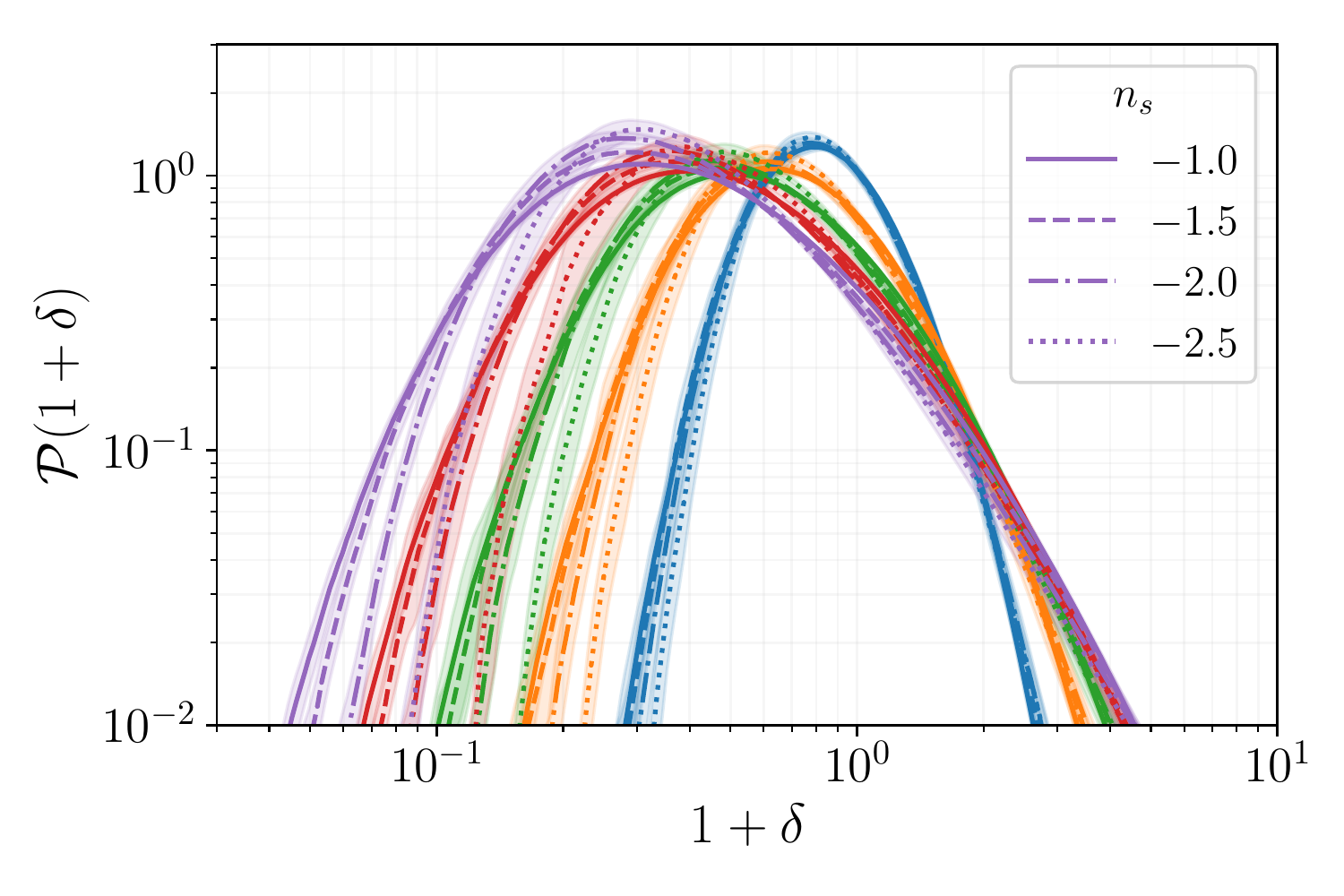}
    \caption{Real-space density PDFs for different $\sigL$ and $n_s$. Blue, orange, green, red and purple lines show $\sigL=0.4, 0.6, 0.8, 1.0,$ and $1.2$, respectively, and $n_s$ is given by line style.}
    \label{fig:rPDF}
\end{figure}

Because the real-space PDF does not explicitly depend on redshift, we combine PDFs at different redshifts with the same $n_s$ and $\sigL$. We report the PDFs as the mean of the jackknife resampled PDFs at these redshifts, and uncertainties as the $\sqrt{7}$ times the standard deviation of them. 
In Figure \ref{fig:rPDF}, we show the real-space PDFs
with different $n_s$ and $\sigL$, with uncertainties shown as the faint bands.

\begin{figure}
    \centering
\includegraphics[width=0.48\textwidth]{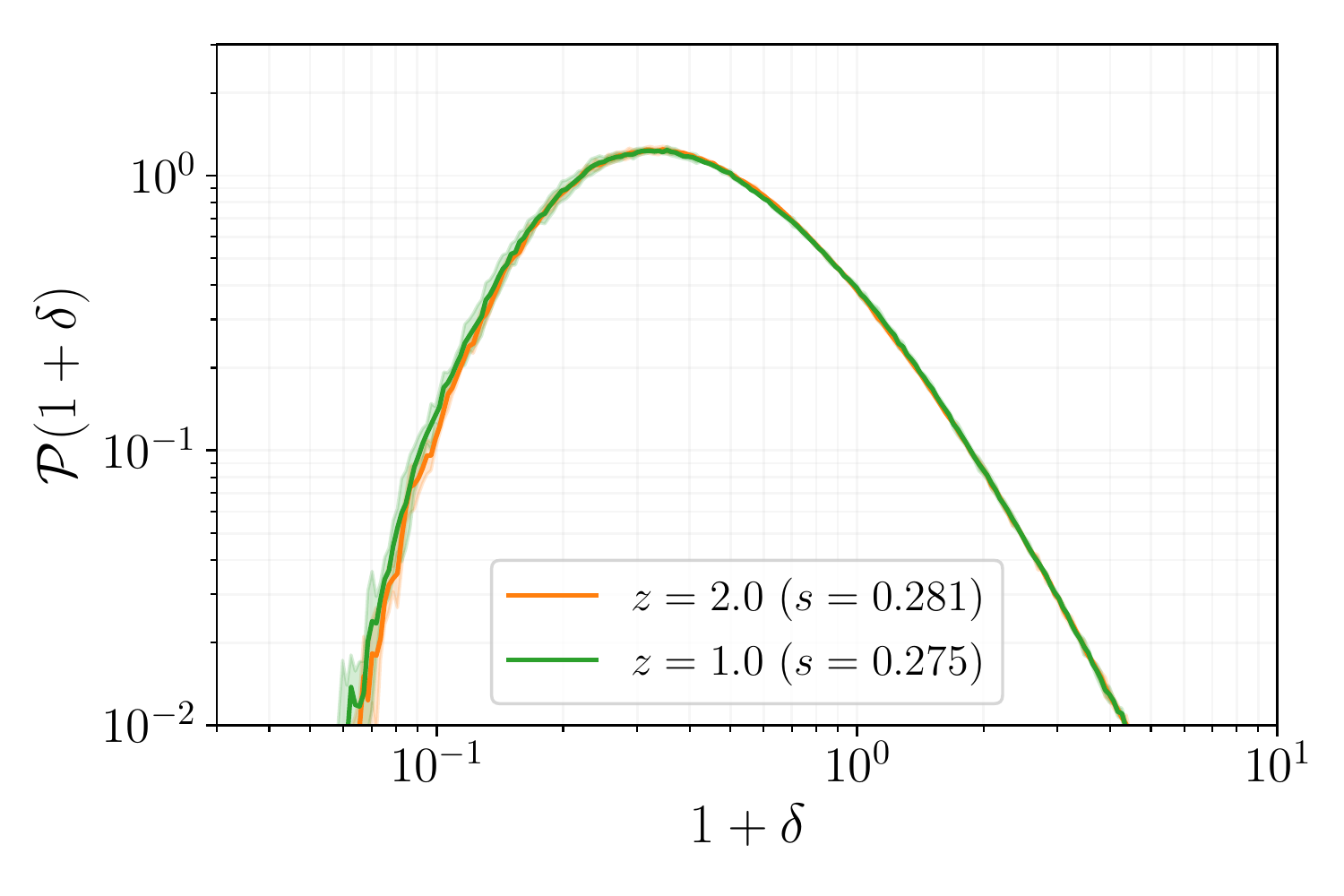}
    \caption{Redshift-space density PDFs for $n_s=-1.5$ and $\sigL=0.8$ measured at redshifts $z=2$ (orange) and $z=1$ (green). Faint bands show the uncertainty due to sample variance. The two curves are almost identical. The distortion parameter $s$ (Equation \ref{eq:s}) calculated directly from the simulation at these two redshifts is very similar.}
    \label{fig:zPDF_scaleV}
\end{figure}

\begin{figure*}
    \centering
    \includegraphics[width=0.45\textwidth]{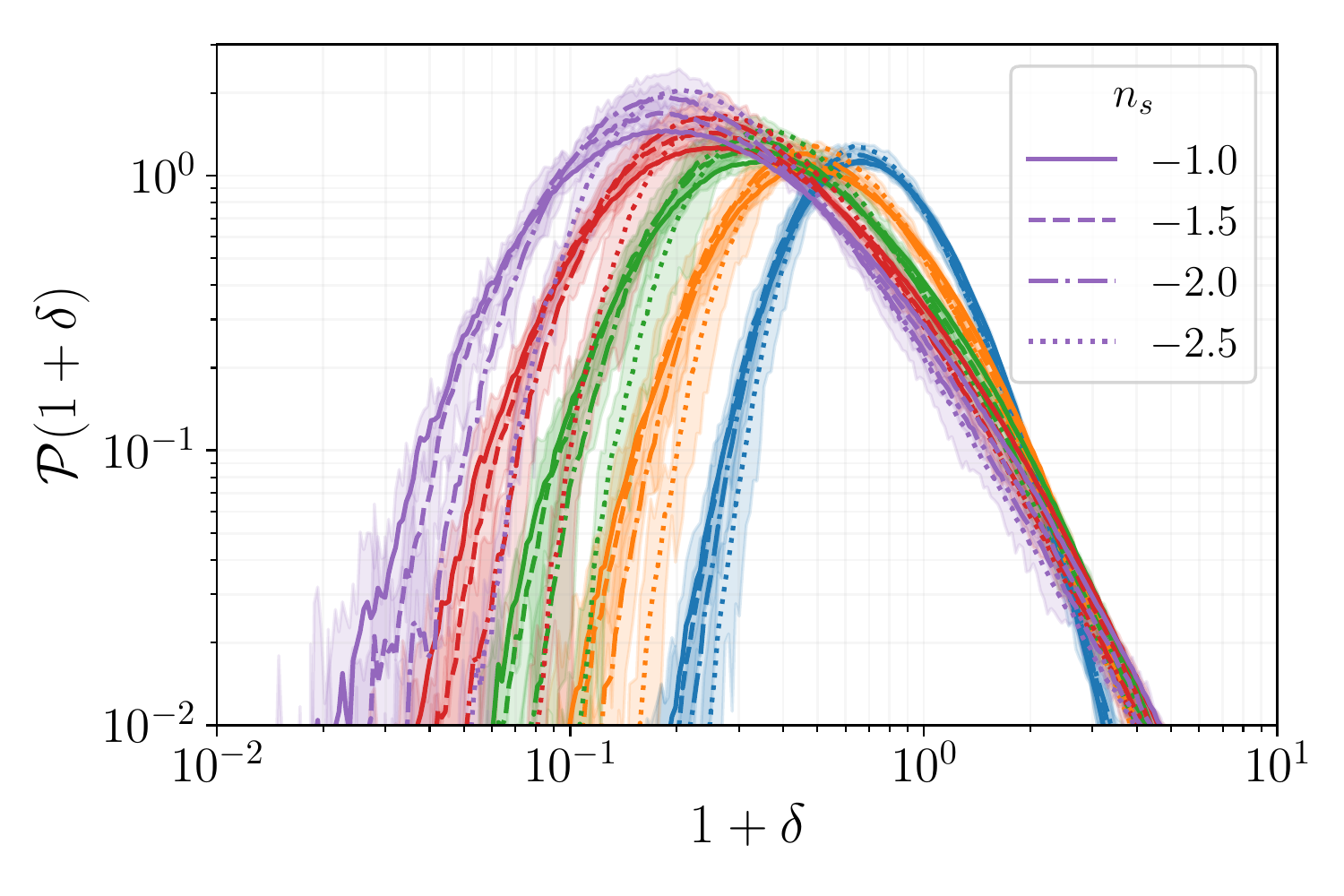}
    \includegraphics[width=0.45\textwidth]{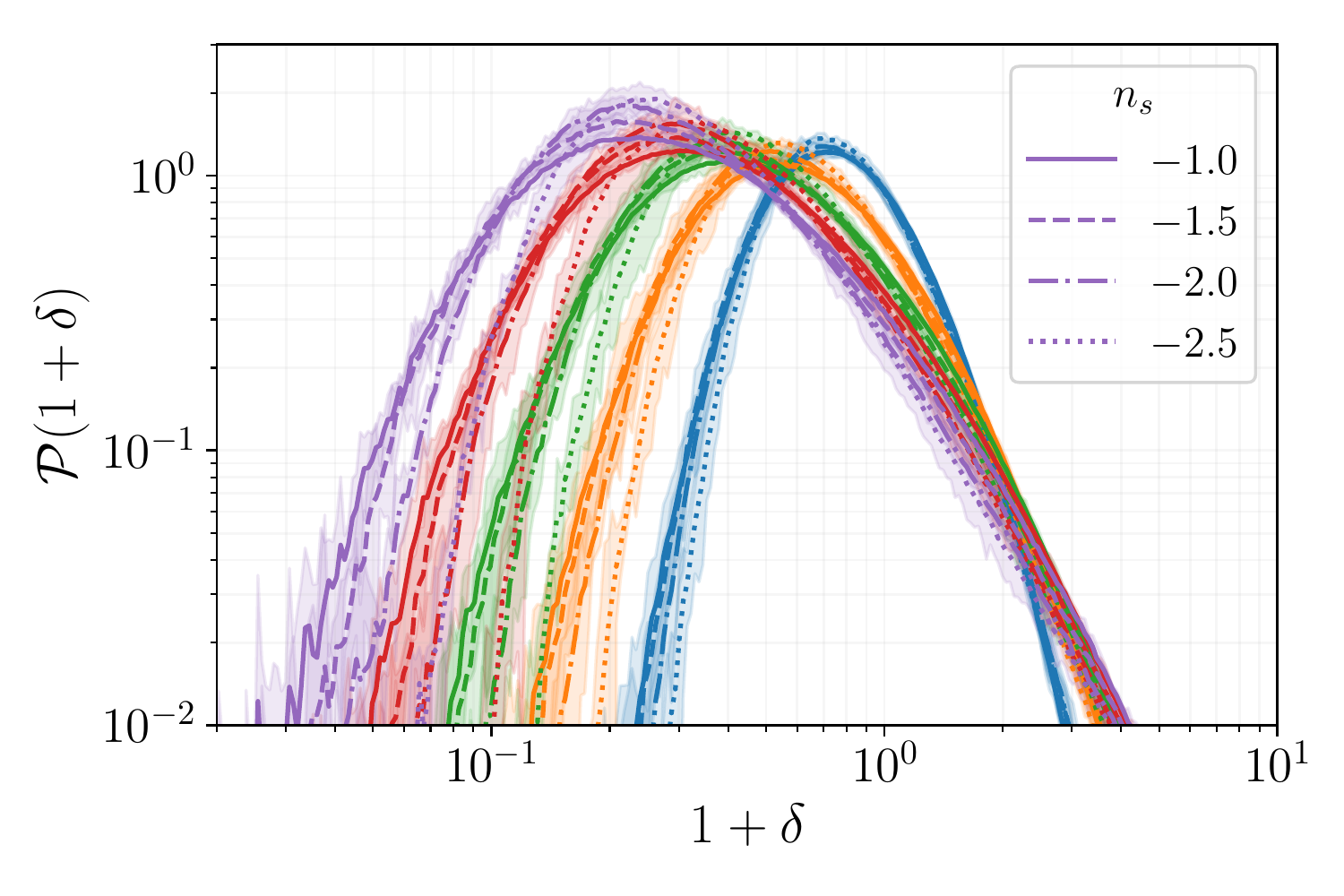}
    \caption{PDFs of Redshift-space density (left) and the geometric mean of real-space density and redshift-space $\sqrt{\Delta_r \Delta_z}$ density (right) for different values of $\sigL$ and $n-s$.}
    \label{fig:zPDF}
\end{figure*}

\begin{figure}
    \centering
    \includegraphics[width=0.45\textwidth]{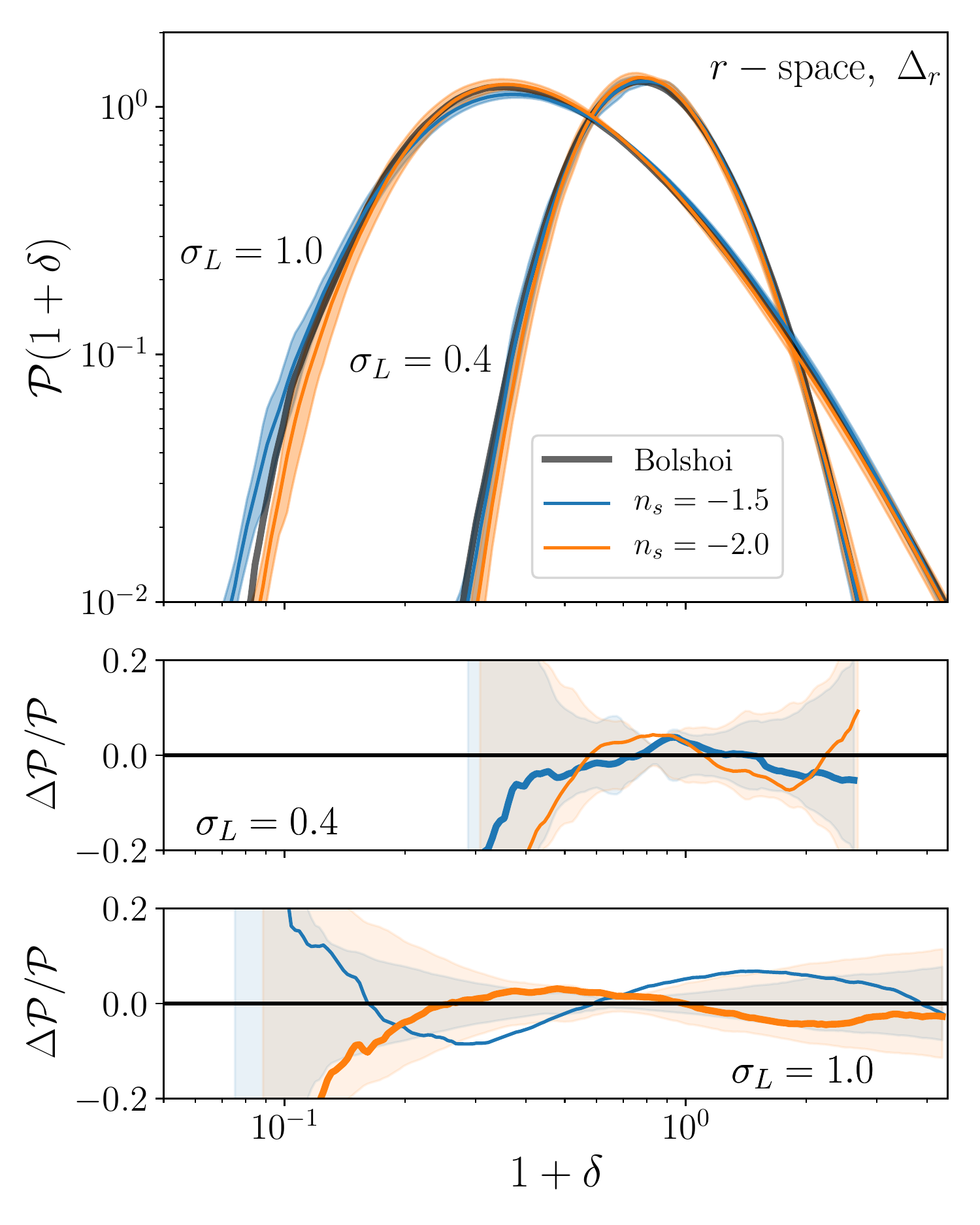}
    \caption{Comparison of real-space PDFs between the Erebos\_CBol simulations and self-similar simulations. In the upper panel, the black curves are the real-space PDF from the Erebos\_CBol simulations for $\sigL=1.0$ (wider one) and $\sigL=0.4$ (narrower one). Blue curves are from the self-similar simulation with $n_s=-1.5$ and orange curves are from $n_s=-2.0$. The middle panel shows the relative differences between the PDF from these self-similar simulations and the Erebos\_CBol simulation for $\sigL=0.4$. The bands show uncertainties due to sample variance. The lower panel is the same as the middle panel except for the value of $\sigL=1.0$.}\label{fig:rs_bol}
\end{figure}

\begin{figure}
    \centering
    \includegraphics[width=0.45\textwidth]{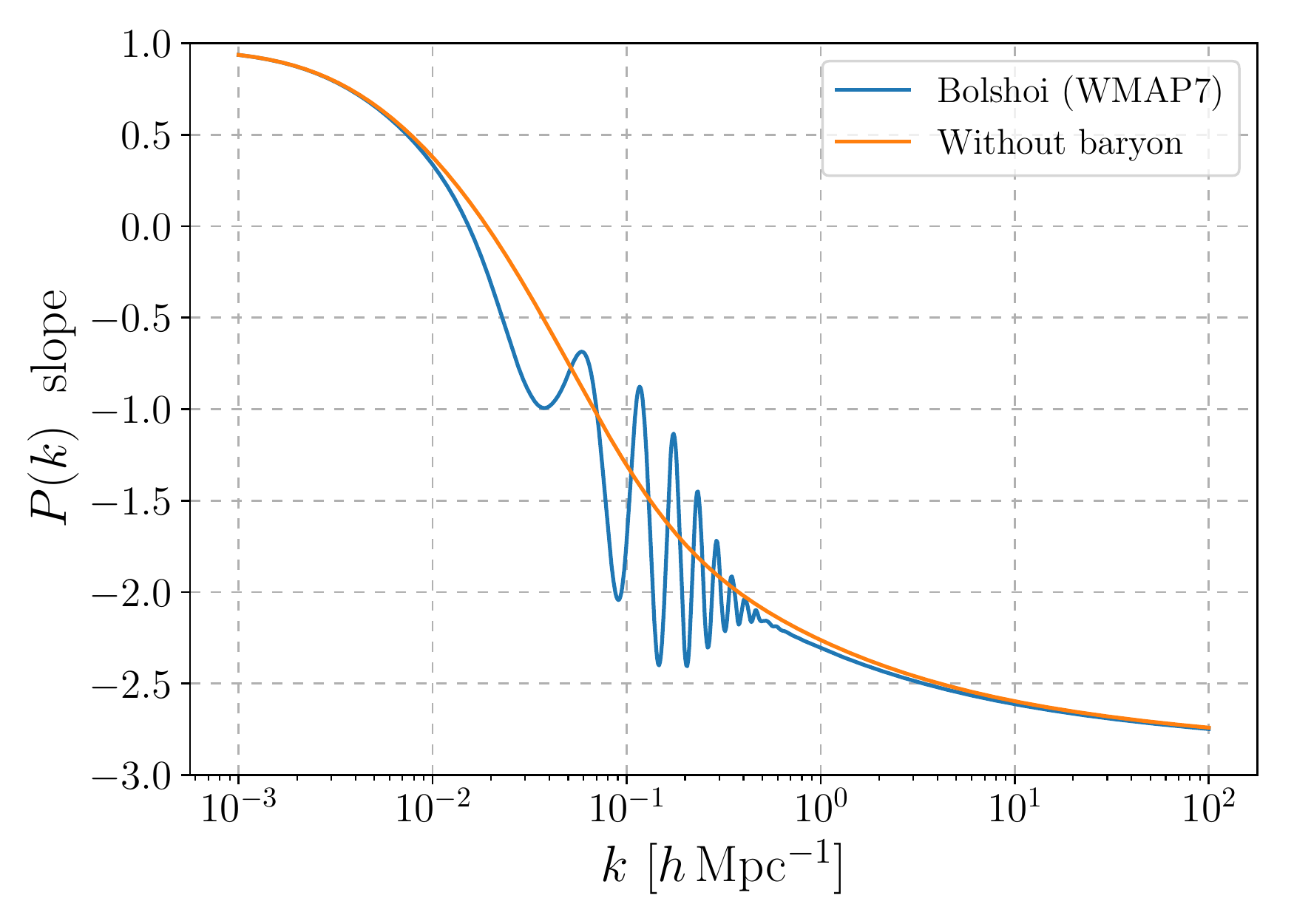}
    \caption{Blue line: the slope of the initial power spectrum used in the Erebos\_CBol simulations. Orange line: the same as the blue one but with dark matter replacing baryons, thus removing baryonic acoustic oscillation. }\label{fig:Pk}
\end{figure}

\begin{figure}
    \centering
    \includegraphics[width=0.45\textwidth]{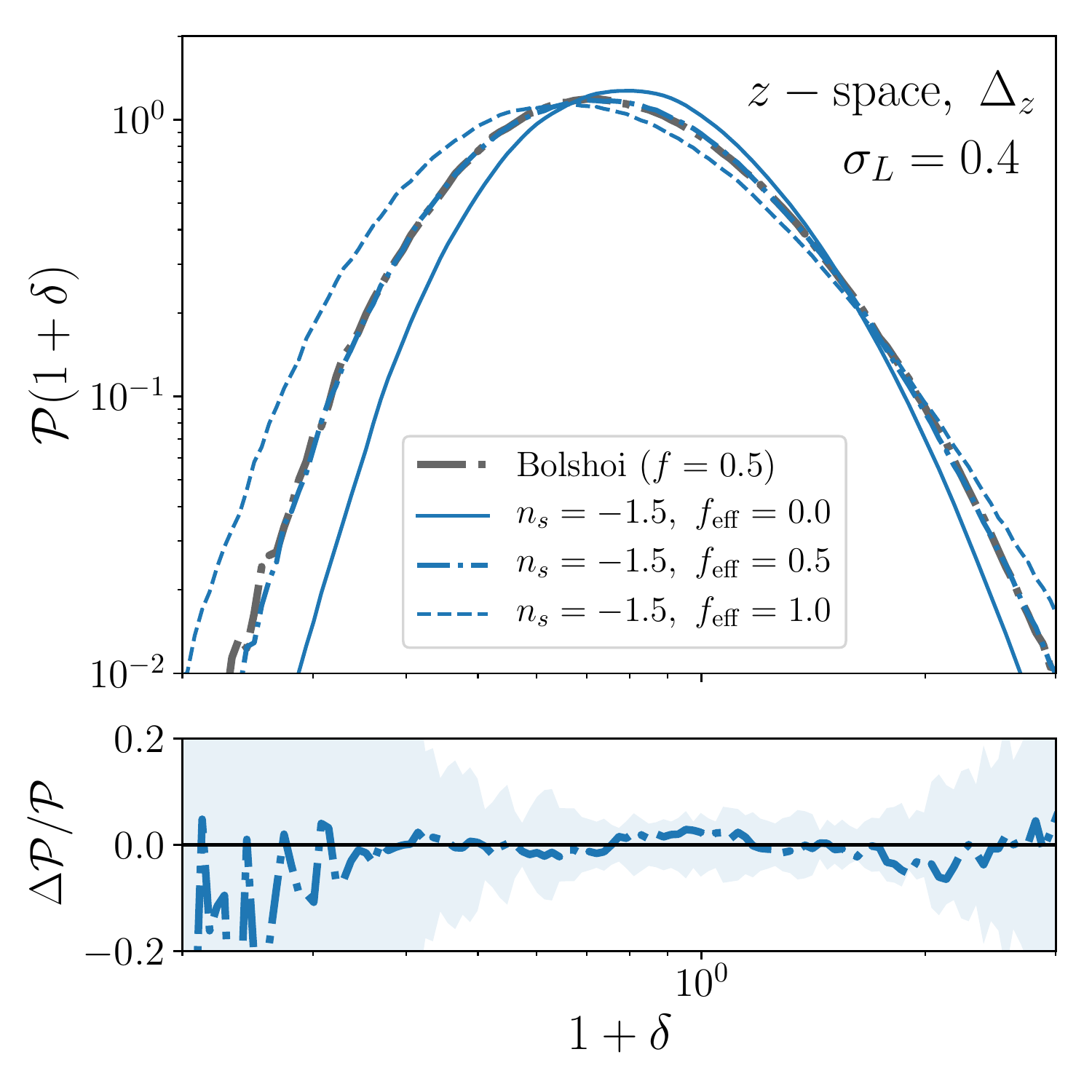}
    \caption{Upper panel: The black dash-dotted curve shows the redshift-space PDF from the Erebos\_CBol\_L1000 simulation at $z=0$. At this redshift, the Bolshoi cosmology has $f(z)\approx0.5$. The blue dashed curve shows the redshift-space PDF from the self-similar simulation with $n_s=-1.5$. The blue dash-dotted curve is the same as the dashed curve except calculated with velocity scaled by $f_{\rm eff}=0.5$. The blue solid curve is the same as the dashed curve except calculated by $f_{\rm eff}=0$ velocity, which is equivalent to the real-space density PDF. 
    All curves are PDFs with $\sigL=0.4$. Lower panel: relative differences between the dash-dotted curves in the upper panel. The band shows uncertainty due to sample variance in the self-similar simulation.}\label{fig:zs_bol}
\end{figure}

\begin{figure}
    \centering
    \includegraphics[width=0.45\textwidth]{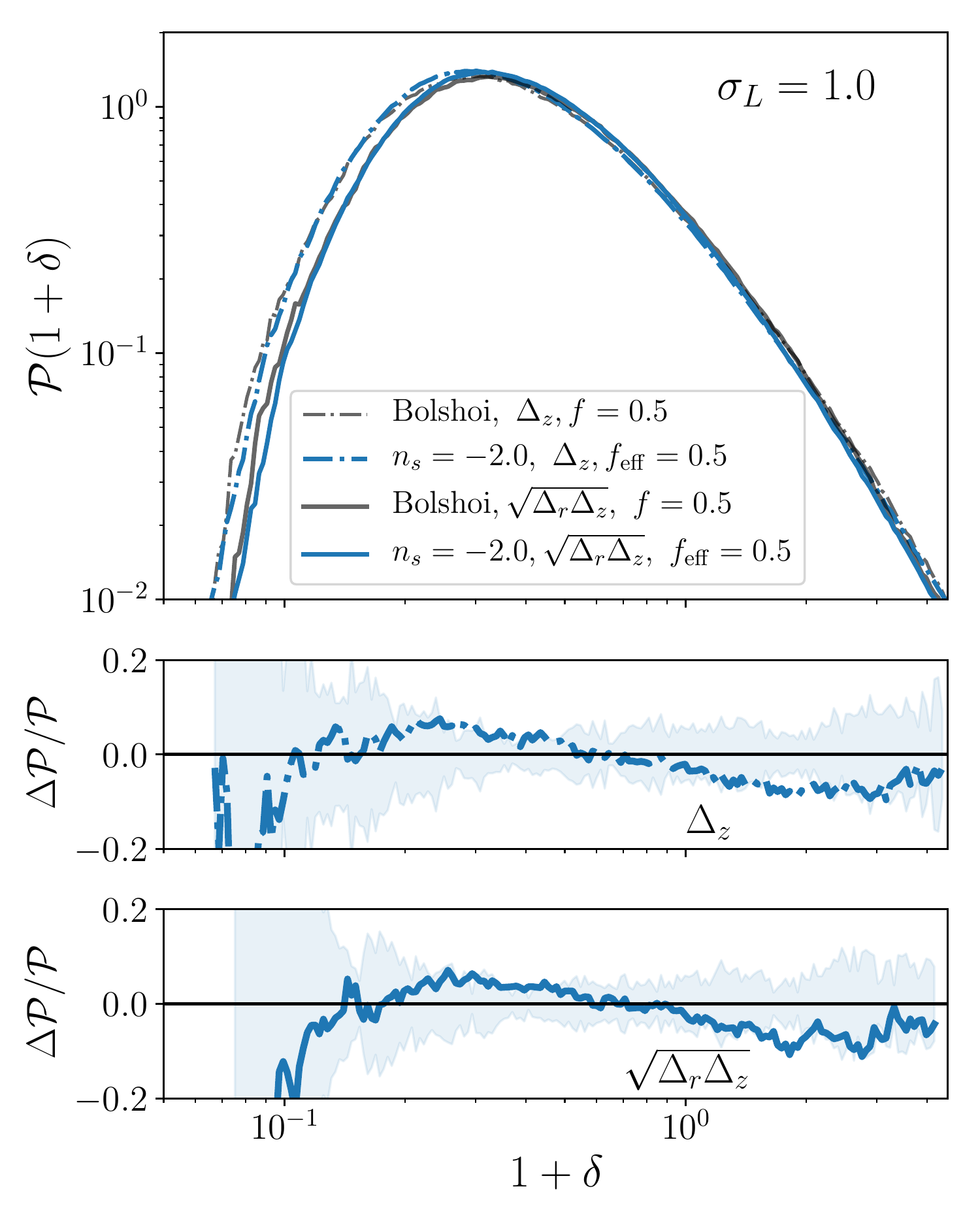}
    \caption{Upper panel: the grey dash-dotted line is the redshift-space PDF from the Erebos\_CBol\_L250   simulation at $z=0$, and the blue dash-dotted line is from self-similar simulation of $n_s=-2.0$, with velocity scaled by $f_{\rm eff}=0.5$. The solid lines are the same except for $\sqrtJD$ PDFs. All lines are for $\sigL=1.0$. Middle and lower panels: relative differences between the blue and black curves, and bands show uncertainty due to sample variance. }\label{fig:zs_sqrtJ_bol}
\end{figure}

\begin{figure}
    \centering
    \includegraphics[width=0.48\textwidth]{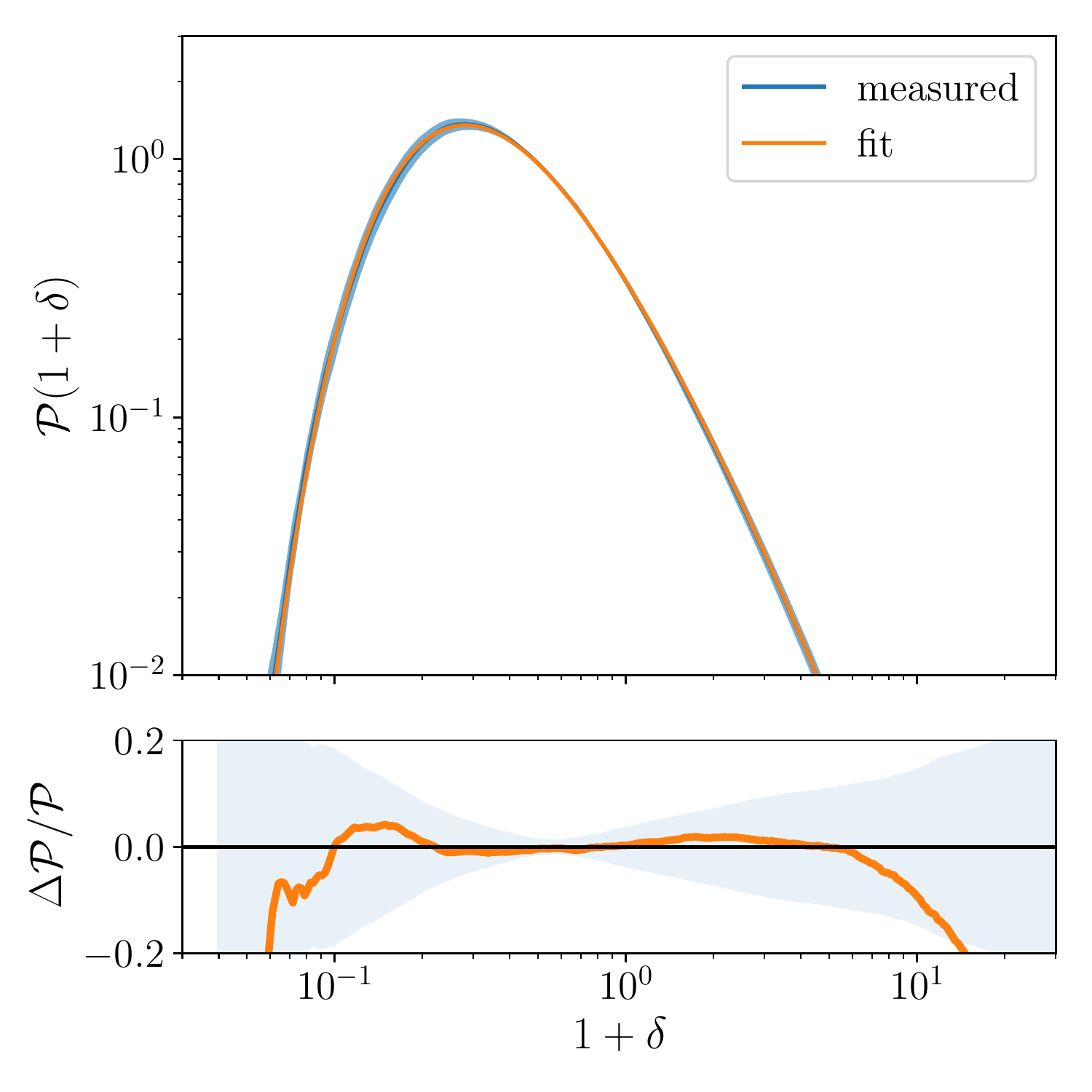}
    \caption{A representative fit from Table \ref{tab:realPDFfitting} to the real-space density PDF with $\sigL=1.2,\ n_s=-2$. For $1+\delta\in[10^{-1},\ 10^1]$, the fit is accurate to $\approx10\%$, with the dominant uncertainty coming purely from sample variance.}
    \label{fig:fit}
\end{figure}

\begin{figure*}
    \centering
    \includegraphics[width=0.98\textwidth]{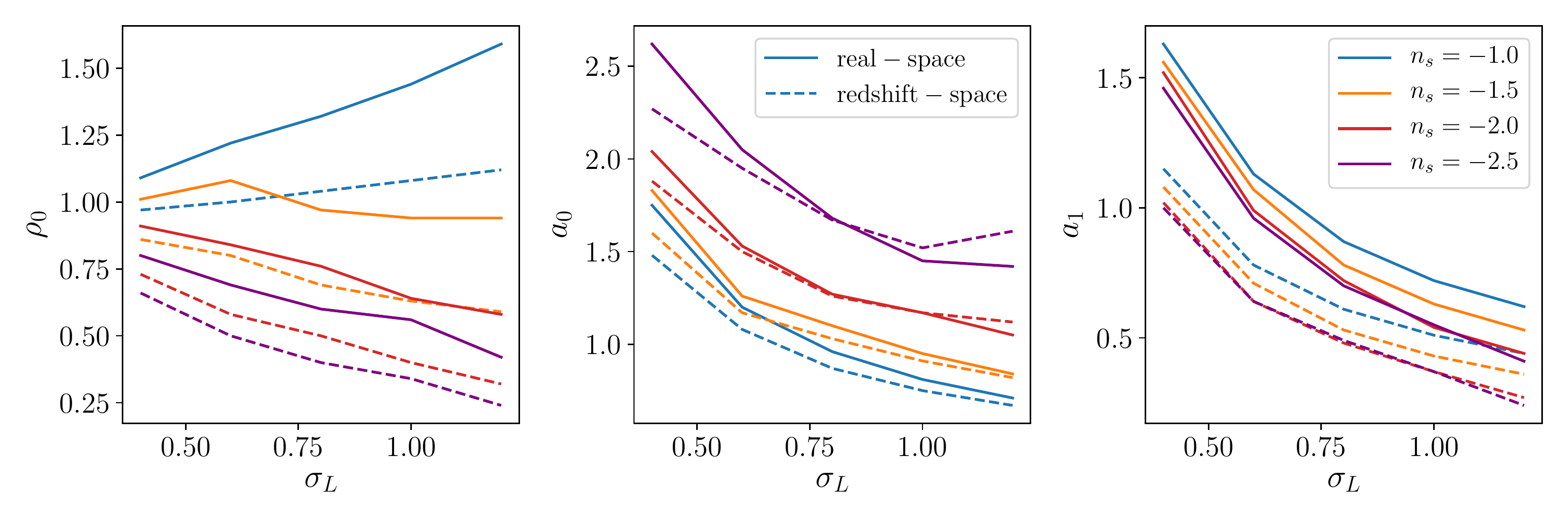}
    \caption{Best-fit $\rho_0$ (left), $a_0$ (middle) and $a_1$(right) as a function of $\sigL$ for different $n_s$. The solid lines are for the real-space density PDF and dashed lines are for the redshift-space density PDF with $f=1$.}
    \label{fig:fitfit}
\end{figure*}

\subsubsection{Redshift-Space Density and $\sqrtJD$ PDFs}

As explained in Section \ref{sec:param_zsPDF}, for a self-similar cosmology the distortion parameter, $s_L$, should be redundant because $f(z)\equiv 1$. In Figure \ref{fig:zPDF_scaleV} we show the redshift-space density PDFs for the same $\sigL$ and $n_s$ but at different redshifts $z=2$ and $z=1$. As expected, the two redshift-space density PDFs are almost the same. We measure the distortion parameter $s$ (Equation \ref{eq:s})
directly from the simulation at these two redshifts and they only differ by $2\%$. This holds true for all the redshift-space density PDFs in the ($n_s, \sigL$) parameter space we study (Table \ref{tab:param_grid}). Thus we can use the distortion parameter calculated in the linear region as the third parameter for redshift-space density PDFs. For $\sqrtJD$, we obtain similar results.

Like the real-space PDFs, to report the PDFs for the redshift-space density and $\sqrtJD$, we combine snapshots with the same $\sigL$ and $n_s$ at different redshifts. We use the jackknife method to estimate the uncertainty. Specifically, for each snapshot and each drawing direction, we create $8$ subsamples by randomly discard $1/8$ of the sightlines. Then we calculate the PDFs for each subsample, and use the $\sqrt{7}$ times standard deviation as the uncertainty.

In Figure \ref{fig:zPDF} we show the redshift-space density and the $\sqrt{\Delta_r  \Delta_z}$ PDFs for different $\sigL$ and $n_s$.

\subsection{$\Lambda$CDM Cosmology}

In the previous sections we have shown that density PDFs in self-similar cosmologies can be well-parameterized with $\sigL$ and $n_s$. As we argued in Section \ref{sec:param_zsPDF}, generic cosmologies should also depend on $f(z)$, but $f(z)\equiv 1$ for power-law cosmologies. It is natural to examine how effective a dependence on all three parameters is at modeling the density PDF of $\Lambda$CDM cosmologies. For a given smoothing kernel, structures on spatial scales much smaller or larger than the smoothing kernel should not significantly impact the shape of the density PDF. Therefore, such a parameterization should provide a reasonable approximation to generic PDFs, provided that $n_s$ is evaluated close to the adopted smoothing scale. With a $3$D Gaussian kernel, it is natural to adopt the power spectra slope at scale $k=1/R$. In this section, we test how well the PDFs derived from self-similar cosmologies approximate the true density PDFs of a simulation with a $\Lambda$CDM cosmology.

We calculate $\Lambda$CDM density PDFs using the Erebos\_CBol simulation suite \citep{diemer2014}. These simulations were run in a WMAP7-like cosmology identical to the one used by the Bolshoi simulation \citep{klypin2011}, with $\Omega_m = 1 - \Omega_\Lambda = 0.27$, $\Omega_b = 0.0469$, $h_{100} = 0.7$, $n_s = 0.95$, and $\sigma_8 = 0.82$. We use the suite's $L=1000\ \CHIMP$ box (Erebos\_CBol\_L1000; $N^3=1024^3$, $m_p=7.0\times10^{10}\ h^{-1}M_\odot$, $\epsilon=33\ h^{-1}{\rm kpc}$) when studying small $\sigma_L$ and the suite's $L=250\ \CHIMP$ box (Erebos\_CBol\_L250; $N^3=1024^3$, $m_p=1.1\times10^{9}\ h^{-1}M_\odot$, $\epsilon=5.8\ h^{-1}{\rm kpc}$) when studying larger $\sigma_L$.

In the upper panel of Figure \ref{fig:rs_bol}, the solid grey lines show the real-space density PDF at $z=0$ for different $\sigL$. The PDF on the right is calculated from Erebos\_CBol\_L1000 with a smoothing scale of $R=9.6\ \CHIMP$, corresponding to $\sigL=0.4$. The solid blue and orange lines show the density PDFs from the self-similar cosmologies with slopes $n_s=-1.5$ and $-2.0$, respectively. The overall shape of both PDFs are very similar to the Erebos\_CBol\_L1000 PDF, particularly the $n_s=-1.5$ PDF. This agreement can be explained by the slope of the matter power spectrum at this smoothing scale. In Figure \ref{fig:Pk}, we show the power spectrum used in the simulation in blue. Due to baryon acoustic oscillations, the power spectrum fluctuates rapidly at $k=0.05\ h\ {\rm Mpc^{-1}}\sim 0.5\ h\ {\rm Mpc^{-1}}$, but these fluctuations should contribute little to the overall density PDF because the amplitude of baryonic acoustic oscillations and thus their contribution to the rms density fluctuation is small. In order to estimate the ``average'' spectral slope at a given scale, we show the power spectrum slope for the same model with $\Omega_b = 0$. In the absence of baryons there are no baryon acoustic oscillations. At $R=9.6 \ \CHIMP$, the $\sigL$ calculated using the orange curve is $0.38$, similar to the value $\sigL=0.4$ from the Erebos\_CBol initial power spectrum. At this scale ($k=1/R = 0.11 {\rm ~~} h\ {\rm Mpc^{-1}}$), the average slope is $-1.4$, similar to the slope of the self-similar cosmology whose PDF approximates the PDF of this simulation the best. The leftmost black PDF was calculated from Erebos\_CBol\_L250 at $z=0$ with a smoothing scale of $R=2.8\ \CHIMP$, corresponding to $\sigL=1.0$. This PDF is more similar to the orange line, e.g., from the self-similar simulation with $n_s=-2.0$, because at $k=1/R = 0.36 {\rm ~~} h\ {\rm Mpc^{-1}}$ the average slope is $-2.0$ (Figure \ref{fig:Pk}).

In Figure \ref{fig:zs_bol}, the grey dash-dotted line shows the redshift-space density PDF from Erebos\_CBol\_L1000 at $z=0$ and smoothed at $R=9.6\ \CHIMP$ ($\sigL=0.4$). We compare it with the redshift-space density PDF (dashed blue line) from the $n_s=-1.5$ self-similar simulation. This is the average slope of the power spectrum in the Bolshoi cosmology at this smoothing scale. They differ significantly. This is expected since their distortion parameter $s_L=f(z) \sigL /\sqrt{5}$ is different --- the self-similar cosmologies have $f(z)=1$, while at $z=0$, the Bolshoi cosmology has $f(z)=0.49$.

Can we use self-similar simulations to approximate the redshift-space PDFs in cosmologies whose distortion parameter $s \propto f(z) \neq 1$? One solution is to artificially scale the velocity fields in the self-similar simulation by a factor of $f_{\rm eff}$, so that the change in velocity dispersion compensates the difference in $f(z)$. We calculate the redshift-space density the same way as before, except using the velocity scaled by a factor of $0.5$ and $0.0$. The redshift-space density PDFs calculated in these ways are shown as the blue dash-dotted lines and the blue solid lines, respectively. As predicted by our model, rescaling by a factor of 0.5 causes the $\Lambda$CDM PDFs to agree well with the ones derived from self-similar simulations with $n_s=-1.5$.

This example demonstrates that using self-similar simulations we can approximate the redshift-space density PDFs in other cosmologies by scaling the velocity field. We test this method on different spatial scales. In Figure \ref{fig:zs_sqrtJ_bol}, we show the redshift-space density PDF from Erebos\_CBol\_L250 at the same redshift ($z=0$) smoothed by $R=2.8\ \CHIMP$ ($\sigL=1.0$) as the grey dash-dotted line. At this scale, the average slope of the linear power spectrum is $n_s \approx -2.0$, and we find that the grey line is well-approximated by the redshift-space PDF from self-similar simulation of $n_s=-2.0$ with velocity scaled by a half. We also check the PDFs of $\sqrt{\Delta_r \Delta_z}$ (solid lines), which also show good agreement. We also test this method at different redshifts $z=0.5$ ($f(z)=0.73$) and $z=1$ ($f(z)=0.85$), and we find similar levels of agreement between Erebos\_CBol and scaled self-similar PDFs.
This test confirms that the set of parameters $\sigL, n_s$, and $f$ is indeed a good way to parameterize the redshift-space density and $\sqrt{\Delta_r \Delta_z}$ PDFs in typical $\Lambda$CDM cosmologies.

\section{Fitting formula}

We have publicly released these density PDFs online\footnote{ \url{https://github.com/huanqing-chen/density\_pdf\_ss\_cosmo}} for a grid of parameters ($\sigL$, $n_s$, $f$) with $\sigL=0.4,0.6,0.8,1.0,1.2$, $n_s=-1.0, -1.5, -2.0, -2.5$ and $f=1, 0.75, 0.5$.
This is the parameter range where we can robustly measure the density PDF to a precision of $\sim 10\%$ for $\mathcal{P}(1+\delta)>0.01$. 

We also provide analytic fits to these PDFs. \citet{klypin2018} showed that the shape of the real-space density PDF can be well approximated by a ``double-exponential" function:
\begin{equation}
    \mathcal{P}(\rho)=A\rho^{-\alpha} {\rm exp}\left[-\left(\frac{\rho_0}{\rho}\right)^{a_0}\right] {\rm exp}\left[-\left(\frac{\rho}{\rho_1}\right)^{a_1}\right].
\end{equation}

This functional form is very flexible with five free parameters. The two exponential terms control the lower and higher tails, while $\alpha$ controls the shape of the PDF around the mean density. We use this functional form to fit the real-space density PDFs. 
For our parameter space, the density PDF does not vary dramatically, and we find that fixing  $\rho_1=1$ and $\alpha=2$ always results in a good fit.
We use \textsc{emcee} \citep{foreman-mackey2013} to find the best-fitting $a_0$, $a_1$ and $\rho_0$ for each $\sigL$ and $n_s$, which we report in Table \ref{tab:realPDFfitting}. We also use the same functional form to fit the PDFs for the redshift-space density and the geometric mean of the real-space and the redshift-space densities, and the results are reported in Table \ref{tab:zsPDFfitting} and  \ref{tab:sqrtJPDFfitting}, respectively.

We check the accuracy for this fitting formula. In the density range where $\mathcal{P}\gtrsim0.1$, the error is $< 10\%$. In Figure \ref{fig:fit} we show a typical fitting result. This is the real-space density PDF for $\sigL=1.2$ and $n_s=-2$. In the lower panel we show the difference between the fit and the measured PDF in orange, while the blue band shows the sample variance of the measured PDF.

In Figure \ref{fig:fitfit} we plot the best-fit $\rho_0, a_0$ and $a_1$ as functions of $\sigL$ for different $n_s$. The trends are well-behaved and can further be interpolated with low-order polynomials to obtain the parameters for $\sigL$ and $n_s$ in between the grid. The same behavior is seen for $\sqrt{\Delta_r \Delta_z}$. Because with data reported in Tables \ref{tab:realPDFfitting}, \ref{tab:zsPDFfitting}, and \ref{tab:sqrtJPDFfitting} one can easily interpolate PDFs, we do not provide fits of best-parameters as functions of $n_s$, $\sigL$, and $f$.

\begin{table*}
\centering
\caption{Best-fit of $\rho_0$, $a_0$ and $a_1$ for real-space density PDFs , fixing $\alpha=2$, $\rho_1=1$.}
\label{tab:realPDFfitting}
\begin{tabular}{|c|c|c|c|c|c|} 
\hline
\diagbox{$n_s$}{$\sigL$} & 0.4            & 0.6            & 0.8            & 1.0            & 1.2             \\ 
\hline
-1.0                                         & 1.09,1.75,1.63 & 1.22,1.20,1.13 & 1.32,0.96,0.87 & 1.44,0.81,0.72 & 1.59,0.71,0.62  \\ 
\hline
-1.5                                         & 1.01,1.83,1.56 & 1.08,1.26,1.07 & 0.97,1.10,0.78 & 0.94,0.95,0.63 & 0.94,0.84,0.53  \\ 
\hline
-2.0                                         & 0.91,2.04,1.52 & 0.84,1.53,0.99 & 0.76,1.27,0.72 & 0.64,1.17,0.54 & 0.58,1.05,0.44  \\ 
\hline
-2.5                                         & 0.80,2.62,1.46 & 0.69,2.05,0.96 & 0.60,1.68,0.70 & 0.56,1.45,0.55 & 0.42,1.42,0.41  \\
\hline
\end{tabular}
\end{table*}

\begin{table*}
\centering
\caption{Best-fit of $\rho_0$, $a_0$ and $a_1$ for redshift-space density PDFs , fixing $\alpha=2$, $\rho_1=1$.}
\label{tab:zsPDFfitting}
\begin{tabular}{|c|c|c|c|c|c|} 
\hline
\diagbox{$n_s$}{$\sigL$} & 0.4            & 0.6            & 0.8            & 1.0            & 1.2             \\ 
\hline
-1.0               & 0.97,1.48,1.15 & 1.00,1.08,0.78 & 1.04,0.87,0.61 & 1.08,0.75,0.51 & 1.12,0.67,0.44  \\ 
\hline
-1.5               & 0.86,1.60,1.08 & 0.80,1.17,0.71 & 0.69,1.03,0.53 & 0.63,0.91,0.43 & 0.59,0.82,0.36  \\ 
\hline
-2.0               & 0.73,1.88,1.02 & 0.58,1.50,0.64 & 0.50,1.26,0.48 & 0.40,1.17,0.37 & 0.32,1.12,0.27  \\ 
\hline
-2.5               & 0.66,2.27,1.00 & 0.50,1.95,0.64 & 0.40,1.67,0.49 & 0.34,1.52,0.37 & 0.24,1.61,0.24  \\
\hline
\end{tabular}
\end{table*}

\begin{table*}
\centering
\caption{Best-fit of $\rho_0$, $a_0$ and $a_1$ for $\sqrtJD$ PDFs , fixing $\alpha=2$, $\rho_1=1$.}
\label{tab:sqrtJPDFfitting}
\begin{tabular}{|c|c|c|c|c|c|} 
\hline
\diagbox{$n_s$}{$\sigL$} & 0.4            & 0.6            & 0.8            & 1.0            & 1.2             \\ 
\hline
-1.0               & 0.94,1.71,1.37 & 0.96,1.21,0.96 & 0.97,0.98,0.75 & 1.01,0.83,0.63 & 1.03,0.74,0.55  \\ 
\hline
-1.5               & 0.89,1.75,1.33 & 0.81,1.31,0.89 & 0.71,1.13,0.68 & 0.65,1.00,0.56 & 0.62,0.88,0.49  \\ 
\hline
-2.0               & 0.78,2.03,1.28 & 0.62,1.64,0.83 & 0.55,1.34,0.64 & 0.44,1.25,0.51 & 0.37,1.16,0.40  \\ 
\hline
-2.5               & 0.70,2.45,1.24 & 0.53,2.22,0.82 & 0.46,1.78,0.68 & 0.38,1.61,0.51 & 0.29,1.61,0.38  \\
\hline
\end{tabular}
\end{table*}

\section{Summary}

We have measured and parameterized density PDFs smoothed with $3$D Gaussian kernels from a suite of self-similar cosmological N-body simulations. 
We find that for self-similar cosmologies, real-space density PDFs can be parameterized with only two parameters, $\sigL$ and $n_s$. We argue that real-space density PDFs of generic cosmologies can be inferred directly from these self-similar PDFs by matching $n_s$ to the slope of the power spectrum at the scale over which the density field is smoothed. We then explicitly demonstrate that this is true for a WMAP7-like cosmology.

To parameterizing the redshift-space density and the geometric mean of the real-space and redshift-space densities PDFs we introduce a third "distortion" parameter, $s_L$.
For self-similar cosmologies, this parameter is redundant because $s\propto f(z)\equiv 1$ (Equation \ref{eq:fz}). For generic cosmologies where $f(z)\neq 1$, we propose that one can can approximate the redshift-space density and the geometric mean of the real-space and redshift-space densities PDFs by scaling the velocity of a self-similar simulations to match the value of $s_L$ in a non-scale-free cosmology. We demonstrate that this procedure works for a WMAP7-like cosmology.

We make our PDFs publicly available and provide analytical fits to them. These fits are accurate to $\sim 10\%$. 

\acknowledgments
This work was supported by the NASA ATP grant NNX17AK65G and NASA FINESST grant NNH19ZDA005K.
This manuscript has been co-authored by Fermi Research Alliance, LLC under Contract No. DE-AC02-07CH11359 with the U.S. Department of Energy, Office of Science, Office of High Energy Physics. 
This project is carried out on the Midway cluster at the University of Chicago Research Computing Center.

\bibliographystyle{apj}
\bibliography{main}

\end{document}